\documentclass[reprint,aps,prL,twocolumn,superscriptaddress]{revtex4-1}

\usepackage{graphicx}
\usepackage{dcolumn}
\usepackage{bm}
\usepackage{amsmath}
\usepackage{braket}
\usepackage[english]{babel}
\usepackage[utf8]{inputenc}
\usepackage[dvipsnames]{xcolor}
\usepackage{soul}
\begin{document}

\title{Universal quantum logic in hot silicon qubits}

\author{L. Petit}
\affiliation{QuTech and Kavli Institute of Nanoscience, Delft University of Technology, PO Box 5046, 2600 GA Delft, The Netherlands}
\author{H. G. J. Eenink}
\affiliation{QuTech and Kavli Institute of Nanoscience, Delft University of Technology, PO Box 5046, 2600 GA Delft, The Netherlands}
\author{M. Russ}
\affiliation{QuTech and Kavli Institute of Nanoscience, Delft University of Technology, PO Box 5046, 2600 GA Delft, The Netherlands}
\author{W. I. L. Lawrie}
\affiliation{QuTech and Kavli Institute of Nanoscience, Delft University of Technology, PO Box 5046, 2600 GA Delft, The Netherlands}
\author{N.~W.~Hendrickx}
\affiliation{QuTech and Kavli Institute of Nanoscience, Delft University of Technology, PO Box 5046, 2600 GA Delft, The Netherlands}
\author{J. S. Clarke}
\affiliation{Components Research, Intel Corporation, 2501 NE Century Blvd, Hillsboro, Oregon 97124, USA}
\author{L. M. K. Vandersypen}
\affiliation{QuTech and Kavli Institute of Nanoscience, Delft University of Technology, PO Box 5046, 2600 GA Delft, The Netherlands}
\author{M. Veldhorst}
\affiliation{QuTech and Kavli Institute of Nanoscience, Delft University of Technology, PO Box 5046, 2600 GA Delft, The Netherlands}


\maketitle
\textbf{
Quantum computation requires many qubits that can be coherently controlled and coupled to each other \cite{ladd2010quantum}. Qubits that are defined using lithographic techniques are often argued to be promising platforms for scalability, since they can be implemented using semiconductor fabrication technology \cite{vandersypen_interfacing_2017, veldhorst_silicon_2017, devoret2013superconducting, neill2018blueprint}. However, leading solid-state approaches function only at temperatures below 100 mK, where cooling power is extremely limited, and this severely impacts the perspective for practical quantum computation. Recent works on spins in silicon have shown steps towards a platform that can be operated at higher temperatures by demonstrating long spin lifetimes \cite{petit2018spin}, gate-based spin readout \cite{urdampilleta2018gate}, and coherent single-spin control \cite{yang2019silicon}, but the crucial two-qubit logic gate has been missing. Here we demonstrate that silicon quantum dots can have sufficient thermal robustness to enable the execution of a universal gate set above one Kelvin. We obtain single-qubit control via electron-spin-resonance (ESR) and readout using Pauli spin blockade. We show individual coherent control of two qubits and measure single-qubit fidelities up to 99.3 \%. We demonstrate tunability of the exchange interaction between the two spins from 0.5 up to 18 MHz and use this to execute coherent two-qubit controlled rotations (CROT). The demonstration of `hot' and universal quantum logic in a semiconductor platform paves the way for quantum integrated circuits hosting the quantum hardware and their control circuitry all on the same chip, providing a scalable approach towards practical quantum information.
}

Spin qubits based on quantum dots are among the most promising candidates for large-scale quantum computation \cite{loss_quantum_1998, zwanenburg_silicon_2013, vandersypen_interfacing_2017}. Quantum coherence can be maintained in these systems for extremely long times \cite{veldhorst_addressable_2014} by using isotopically enriched silicon (\textsuperscript{28}Si) as the host material \cite{itoh_isotope_2014}. This has enabled the demonstration of single-qubit control with fidelities exceeding 99.9\% \cite{yoneda_quantum-dot_2017, yang2018silicon} and the execution of two-qubit logic \cite{veldhorst_two-qubit_2015, zajac_resonantly_2018, watson_programmable_2018, huang2019fidelity}. The potential to build larger systems with quantum dots manifests in the ability to deterministically engineer and optimize qubit locations and interactions using a technology that greatly resembles today's complementary metal-oxide semiconductor (CMOS) manufacturing. Nonetheless, quantum error correction schemes predict that millions to billions of qubits will be needed for practical quantum information \cite{fowler2012surface}. Considering that today's devices make use of more than one terminal per qubit \cite{franke2019rent}, wiring up such large systems remains a formidable task. In order to avoid an interconnect bottleneck, quantum integrated circuits hosting the qubits and their electronic control on the same chip have been proposed \cite{veldhorst_silicon_2017, vandersypen_interfacing_2017, li_crossbar_2017}. While these architectures provide an elegant way to increase the qubit count to large numbers by leveraging the success of classical integrated circuits, a key question is whether the qubits will be robust against the thermal noise imposed by the power dissipation of the electronics. Demonstrating a universal gate set at elevated temperatures would therefore be a milestone in the effort towards scalable quantum systems.

Here, we solve this challenge and combine initialization, readout, single-qubit rotations and two-qubit gates, to demonstrate full two-qubit logic in a quantum circuit operating at 1.1 Kelvin. We furthermore examine the temperature dependence of the quantum coherence which we find, unlike the relaxation process \cite{petit2018spin}, to be hardly affected in a temperature range up to one Kelvin.

\begin{figure*}%
	\includegraphics[width=\linewidth]{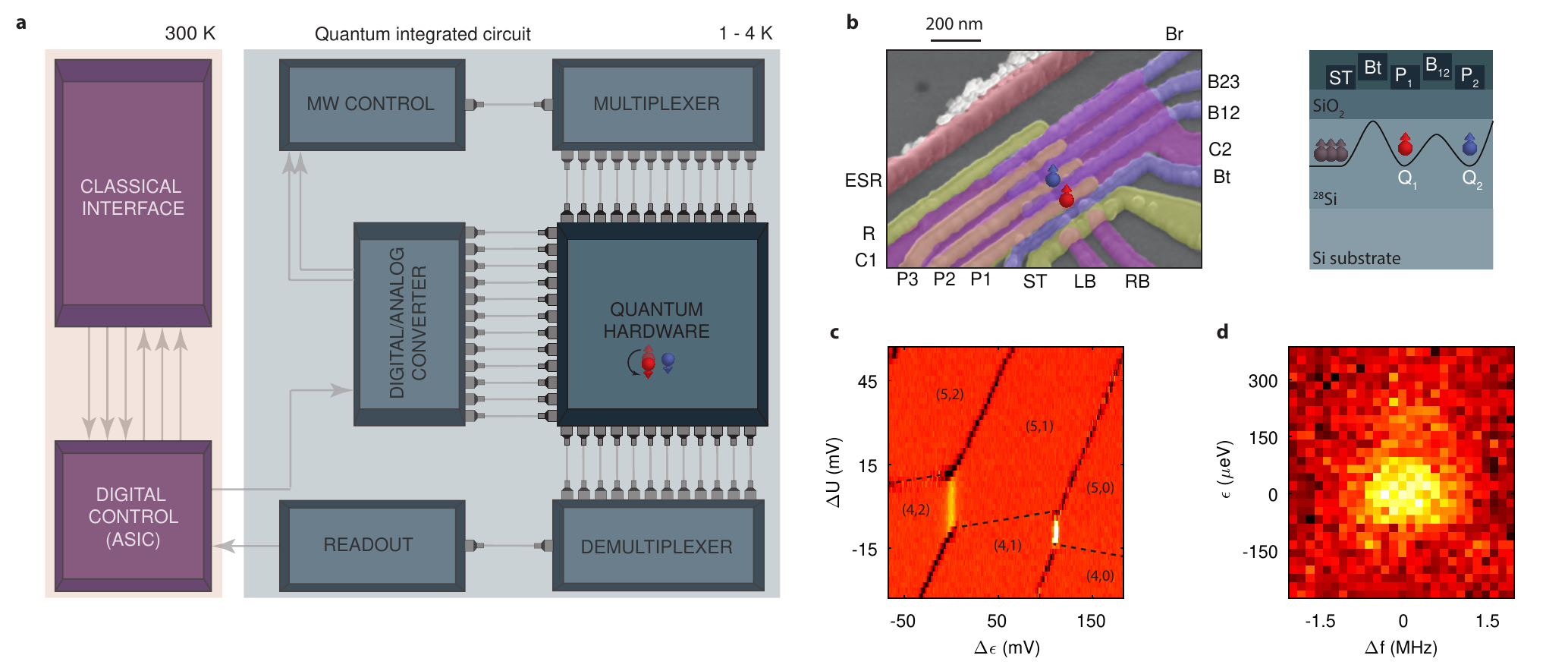}%
	\caption{\textbf{Large-scale approach for silicon qubits} 
	\textbf{a} Quantum integrated circuit as a scalable approach for quantum computing, where the qubits and their control electronics are defined on the same chip. The control functionality that can be integrated is strongly dependent on the available cooling power. When the qubits can be coherent controlled above one Kelvin a broad range of electronics may be integrated, such that communication between room temperature and the coldest stage is limited to only digital signals. 
	\textbf{b} Scanning Electron Microscope (SEM) image of a quantum device identical to the one measured. Gates P1 and P2 define the two quantum dots and the gates Br and B12 control the tunneling rate to the reservoir R and the inter-dot tunnel coupling, respectively. The SET is defined by the top gate ST and the two barriers RB and LB, and it is used both as charge sensor and as reservoir \cite{morello_single-shot_2010}, while the tunnel rate is controlled by Bt. The gates C1 and C2 confine the electrons in the three quantum dots. Gates R, Br, P3 and B23 are kept grounded during the experiment.
	\textbf{c} Electron occupancy as a function of detuning energy between the two quantum dots $\epsilon$ and on-site repulsion energy $U$. The data have been centered at the (4,2)-(5,1) anticrossing. The electron transitions have been measured via a lockin technique, by applying an excitation of 133 Hz on gate B12. Both electrons are loaded from the SET, with Q2 having a tunneling rate significantly lower than Q1. 
	\textbf{d} Readout visibility as a function of $\epsilon$ and microwave frequency applied to Q2. When the readout level is positioned between the singlet-triplet energy splitting and the microwave frequency matches the resonance frequency of Q2, we correctly read out the transition from the state $\ket{\downarrow \uparrow}$ to the blocked state $\ket{\uparrow \uparrow}$. 
	}
\label{fig:Intro}
\end{figure*}

Figure \ref{fig:Intro}a conceptually displays a quantum integrated circuit. Inspired by their classical counterpart where only a few control lines are needed to interact with billions of transistors, a quantum integrated circuit hosts the quantum hardware and its electronic control on the same chip to provide a scalable solution \cite{franke2019rent}. Here we focus on the quantum hardware of such a circuit, which we implement using silicon quantum dots.

Figure \ref{fig:Intro}b shows the silicon quantum dot device. The qubits are realized in an isotopically purified \textsuperscript{28}Si epi-layer with a \textsuperscript{29}Si residual concentration of 800 ppm. The fabrication of the quantum dot device is based on an overlapping gate-scheme to allow for tightly confined quantum dots \cite{lawrie2019quantum}. Electrons can be loaded either from the reservoir or from the single-electron-transistor (SET) \cite{morello_single-shot_2010}, which is also used for charge sensing. To allow for coherent control over the electron spins, AC currents are applied through the on-chip aluminum microwave antenna.

Figure \ref{fig:Intro}c shows a charge stability diagram of the double quantum dot, where the qubits Q1 and Q2 and their coupling are defined by using the gates P1, B12, and P2. Since we can freely choose the occupancy of the two quantum dots we tune to the regime where we obtain optimal exchange coupling, which we find with one and five electrons for Q1 and Q2 respectively. We then operate the system close to the (5,1)-(4,2) charge anticrossing.

Single spins are often initialized via energy-selective tunneling to a nearby reservoir \cite{elzerman2004single}. However, this method requires a Zeeman splitting much higher than the thermal broadening, limiting the fidelity and making the method unpractical for high temperature operation. Instead, Pauli spin blockade offers a convenient mechanism to perform initialization and readout \cite{vandersypen_interfacing_2017, urdampilleta2018gate}, with a relevant energy scale corresponding to the singlet-triplet energy splitting, which is set by the large and tunable valley splitting energy in silicon metal-oxide-semiconductor (SiMOS) devices \cite{yang_spin-valley_2013}. This method is more robust against thermal noise and enables independent optimization of the qubit operation frequency. We choose to set the magnetic field to $B$ = 0.25 T, which corresponds to addressable qubits with Larmor frequencies $\nu_{Q1}$ = 6.949 GHz and $\nu_{Q2}$ =  6.958 GHz in the absence of exchange interaction. This low frequency operation reduces the qubit sensitivity to electrical noise that couples in via the spin-orbit coupling \cite{ruskov2018electron}. Additionally it also simplifies the demands on the electronic control circuits and reduces the cable losses.

\begin{figure*}%
	\includegraphics[width=\linewidth]{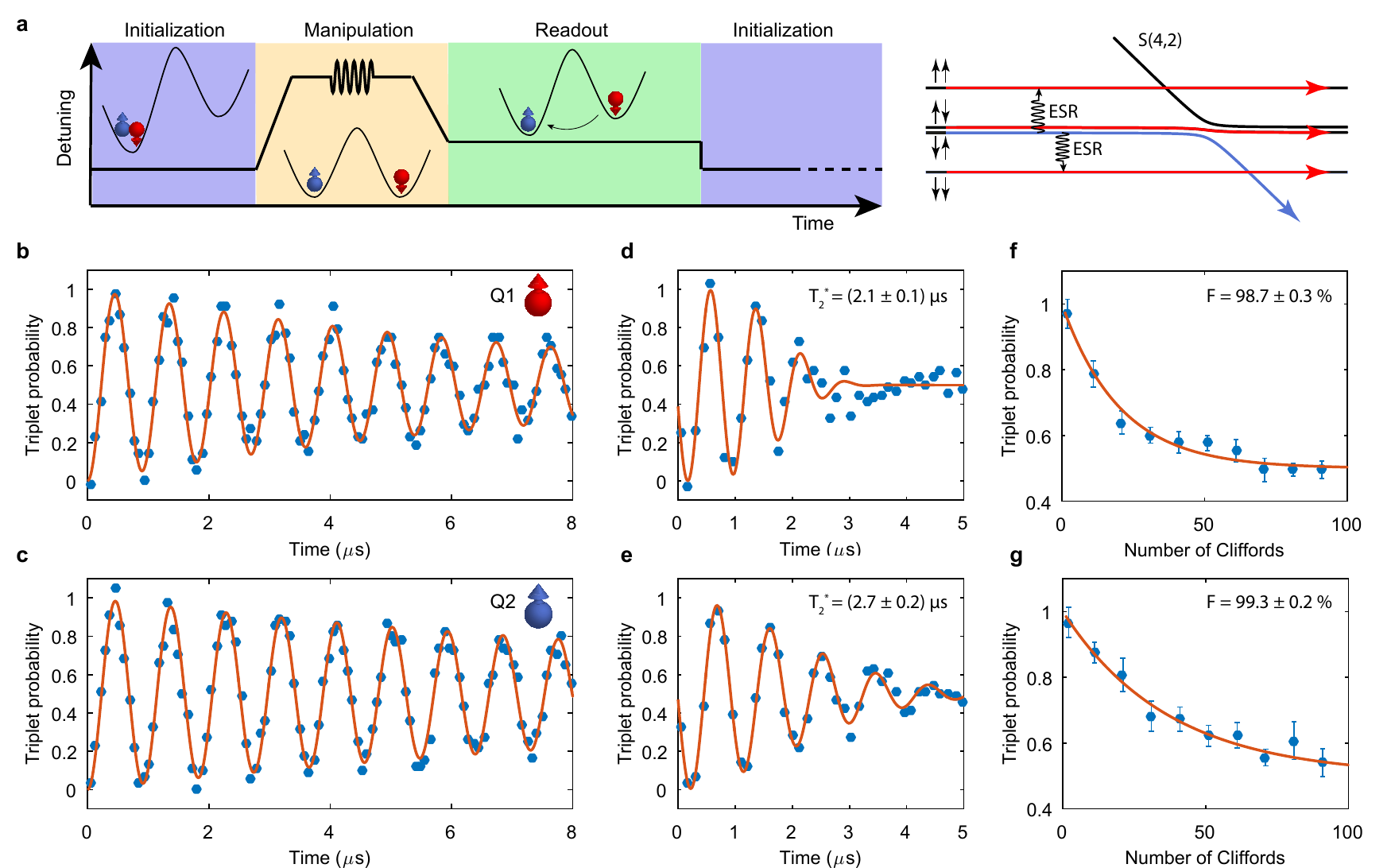}%
    \caption{\textbf{Single-qubit characterization at 1.1 K.} 
    \textbf{a} Pulse sequence used for the experiments. Qubits Q1 and Q2 are defined on the spin states of single electrons, the remaining four electrons in Q2 fill the first levels and do not contribute to the experiment. A voltage ramp allows adiabatic transitions between the (5,1) and (4,2) charge states. Each measurement cycle consists of two of these sequences. The second cycle contains no pulses and it is used as a reference cycle to cancel low-frequency drifts during readout. 
    \textbf{b-c} Rabi oscillations for both qubits as a function of the microwave pulse duration. We extract decay time constants $T_{2(Q1)}^{Rabi} = 8 $ $\mu$s and $T_{2(Q2)}^{Rabi} = 14$ $\mu$s. 
    \textbf{d-e} Decay of the Ramsey fringes for both qubits. The data correspond to the average of four traces where each point is obtained from 500 single-shot traces.
    \textbf{f-g} Randomized benchmarking of the single-qubit gates for both qubits. Each data point is obtained from 500 averages of 20 Clifford sequences, for a total of 10.000 single-shot traces. The fidelity reported refers to the primitive gates, while a Clifford-gate contains on average 1.875 primitive gates. We have normalized the triplet probability to remove the readout errors. 
    }
\label{fig:SingleQubits}
\end{figure*}

The pulse sequence used in the experiment is schematically shown in Fig. \ref{fig:SingleQubits}a. The sequence starts by pulsing deep into the (4,2) charge state, where the spins quickly relax to the singlet state. An adiabatic pulse to the (5,1) regime is applied to initialize the system in the $\ket{\downarrow\uparrow}$ state. At this position in detuning energy $\epsilon$, single- and two-qubit gate operations are performed by applying a microwave burst with variable frequency and duration. The sequence ends by adiabatically pulsing to the anticrossing where readout is performed. The antiparallel spin state with the lowest energy (which is in this experiment the state $\ket{\downarrow \uparrow}$ state) couples directly to the singlet (4,2) charge state. Instead, the remaining antiparallel spin state ($\ket{\uparrow \downarrow}$) and the two parallel spin states ($\ket{\uparrow \uparrow}$, $\ket{\downarrow \downarrow}$) couple to the three triplet (4,2) charge states. This allows to map spin information (singlet or triplet states) to charge configurations ((4,2) or (5,1) states), which can be read out using the SET. As shown in Fig. \ref{fig:Intro}d, the optimal readout position can be obtained by sweeping $\epsilon$ and applying a $\pi$-pulse to Q2. From the detuning lever arm of $\alpha_\epsilon = 0.044$ eV/V, extracted from the thermal broadening of the polarization line, we find a readout window of 155 $\mu$eV where we can efficiently discriminate between the singlet and triplet states.

In this high temperature operation mode, the readout visibility is mainly limited by the broadening of the SET peaks. In order to maximize our sensitivity we subtract a reference signal from each trace, then we average and normalize the resulting traces (for more details on the readout see supplemental material section I). 

Figure \ref{fig:SingleQubits} shows the single-qubit characterization of the two-qubit system. We observe clear Rabi oscillations for both qubits (Fig. \ref{fig:SingleQubits}b, c) as a function of the microwave burst duration. From the decay of the Ramsey fringes (Fig. \ref{fig:SingleQubits}d, e) we extract dephasing times $T_{2(Q1)}^* = 2.1$ $\mu$s and $T_{2(Q2)}^* = 2.7$ $\mu$s, comparable to experiments at similar high temperature \cite{yang2019silicon} and still longer than dephasing times for natural silicon at mK temperatures \cite{watson_programmable_2018, zajac_resonantly_2018}. 

We characterize the performance of the single-qubit gates of the two qubits by performing randomized benchmarking \cite{magesan2011scalable}. In the manipulation phase we apply sequences of random gates extracted from the Clifford group, followed by a recovery gate that brings the system to a triplet state. By fitting the decay of the readout signal as a function of the number of applied gates to an exponential decay we extract qubit fidelities $F_{Q1} = 98.7 \pm 0.3 $ $\%$ and $F_{Q2} = 99.3 \pm 0.2$ $\%$. 

\begin{figure*}%
	\includegraphics[width=\linewidth]{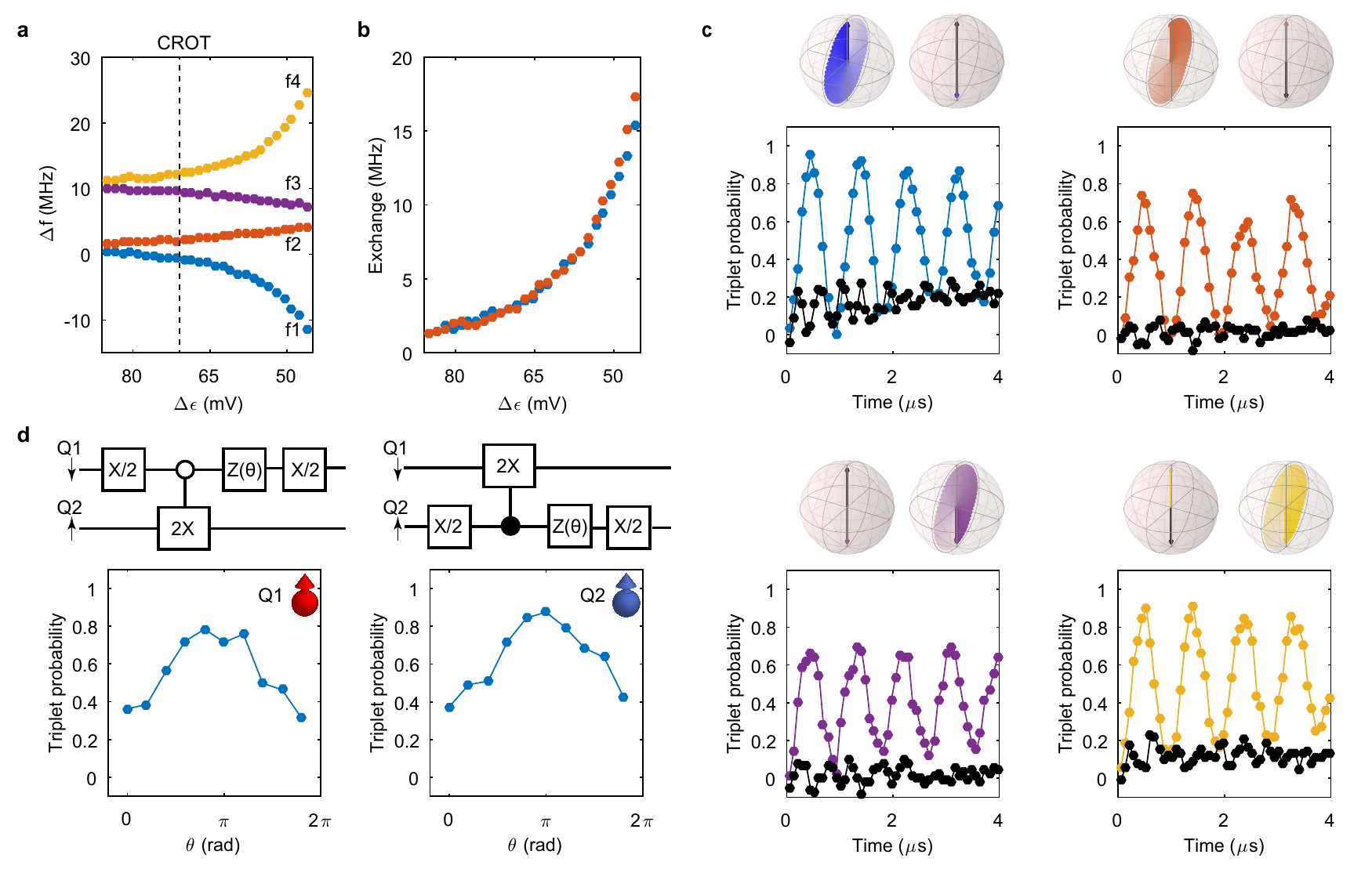}%
	\caption{\textbf{Exchange and two-qubit logic at 1.1 K.}
	\textbf{a} Full exchange diagram obtained from a gaussian fit of the data shown in supplemental material section II. The frequency offset is 6.948 GHz.
	\textbf{b} Exchange energy measured as a function of detuning. The data correspond to $f_2-f_1$ and $f_4-f_3$ as obtained from \textbf{a}.	
	\textbf{c} Conditional rotations on all the frequencies $f_i$, the color code refers to the plot in \textbf{a}. The black lines correspond to the same transition $f_i$, driven with the control qubit in the opposite state. 
	An initialization $\pi$-pulse and recovery $\pi$-pulse is applied to the control qubit for the sequences where either Q1 is in the spin down state or Q2 is in the spin up state. All Rabi frequencies are set to approximately 1 MHz by adjusting the power of the microwave source to compensate for the frequency dependent attenuation of the fridge line.
	\textbf{d} Phase acquired by the control qubit during a CROT operation. A CROT gate, together with a Z-rotation of $\pi/2$ on the control qubit is equivalent to a CNOT operation. $Z$ gates are implemented by a software change of the reference frame. We have normalized the triplet probability to remove the readout errors. 
	}
\label{fig:CROT}
\end{figure*}

We now turn to the two-qubit gate characterization. The ability to tune the exchange interaction \cite{loss_quantum_1998} is the basis to perform two-qubit operations with electrons in quantum dots. By turning on the exchange interaction, either by controlling the detuning energy or the tunnel coupling, the resonance frequencies of each qubit shift depending on the spin state of the other qubit. Figure \ref{fig:CROT}a shows this frequency shift for both qubits as a function of the detuning energy between the two quantum dots, with and without a $\pi$-pulse applied to flip the spin state of the other qubit. The full exchange spectrum is composed of the transitions $f_1$ ($\ket{\downarrow \uparrow} \longrightarrow \ket{\downarrow \downarrow}$), $f_2$ ($\ket{\uparrow \uparrow} \longrightarrow \ket{\uparrow \downarrow}$), $f_3$ ($\ket{\downarrow \downarrow} \longrightarrow \ket{\uparrow \downarrow}$) and $f_4$ ($\ket{\downarrow \uparrow} \longrightarrow \ket{\uparrow \uparrow}$). The exchange interaction, extracted as the differences $f_2-f_1$ and $f_4-f_3$, is plotted in Fig. \ref{fig:CROT}b. We measure tunable $J$ in the range 0.5 - 18 MHz. At even larger exchange couplings the readout visibility drastically reduces, which we attribute to a decrease of $T_{2}^*$ (see Fig \ref{fig:temp} (a)). By fitting the exchange spectrum (see supplemental material section III) we extract a tunnel coupling $t_c$ = 0.8 GHz and a Zeeman energy difference $\delta E_z$= 9.1 MHz. 

Having demonstrated the tunability of the exchange interaction, we use this to demonstrate two-qubit operation. When the exchange is turned on, the resulting shift in resonance frequency can be used to implement state selective ESR transitions (CROT), which are equivalent to a CNOT gate up to single-qubit phases. Figure \ref{fig:CROT}c shows controlled oscillations for both qubits, with the control qubit set either to the spin down or spin up state, where we have set the exchange interaction to $J$ = 2.5 MHz. When we prepare the state of the control qubit such that the target qubit is in resonance with the external microwave control, we observe clear oscillations of the target qubit as a function of the microwave burst duration, with no significant decay after multiple rotations. When we flip the state of the control qubit, the resonance frequency of the target qubit is shifted and the target qubit is not driven by the microwave control. 

In order to investigate the coherence of the two-qubit logic, we apply a sequence where we interleave a CROT operation with duration $2\pi$ in between two $\pi/2$ single-qubit gates applied to the control qubit with variable phase $\theta$. As shown in Fig. \ref{fig:CROT}d, when we invert the second $\pi/2$ pulse ($\theta = \pi$) this cancels out the $\pi$ phase left by the CROT operation on the control qubit and we correctly measure transitions to the $\ket{\downarrow \downarrow}$ and $\ket{\uparrow \uparrow}$ states. This demonstrates the execution of a coherent CROT, since the control qubit maintains its coherence even when the target qubit is driven.

\begin{figure*}%
\includegraphics[width=\linewidth]{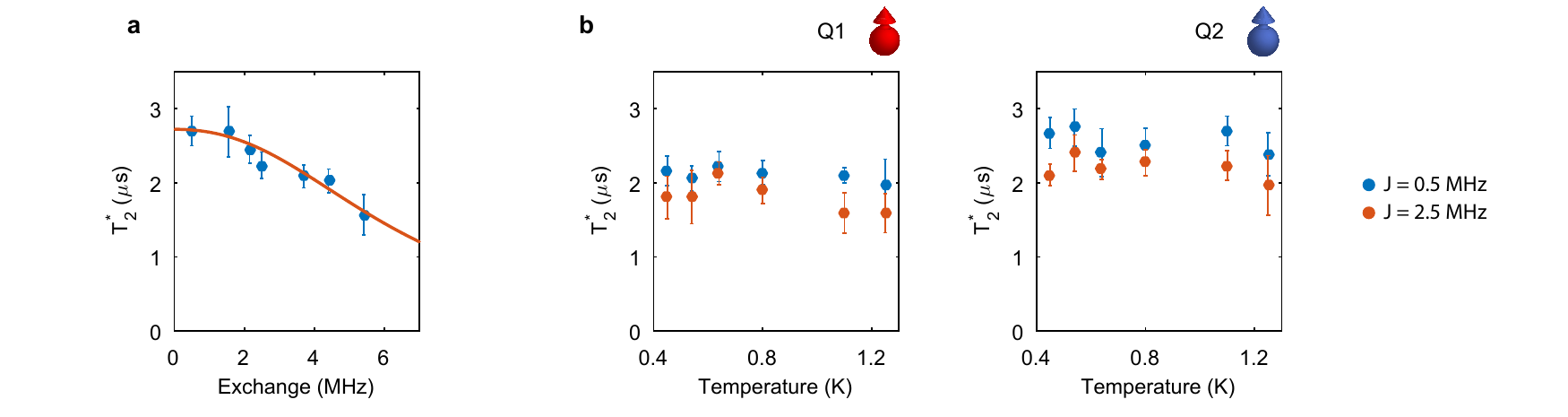}%
	\caption{\textbf{Dephasing dependence on temperature and exchange interaction.}
	\textbf{a} Dephasing time of Q2 as a function of the exchange interaction, fitted with a model taking into account gaussian quasi static noise (see supplemental material section IV).  
	\textbf{b} Temperature dependence of the dephasing time with the exhange interaction set to the minimum obtained by sweeping $\epsilon$ ($J$ = 0.5 MHz) and with the exchange interaction set to acquire the CROT operations of Fig. \ref{fig:CROT} ($J$ = 2.5 MHz).
	}
\label{fig:temp}
\end{figure*}

To further investigate the quantum coherence of the system we measure the decay of the Ramsey fringes for different values of the exchange interaction, see Fig. \ref{fig:temp}a. We find that by increasing the exchange interaction the coherence is reduced, which we explain by the increased qubit sensitivity to electrical noise. We can fit the data with a model (see supplemental material section IV) that includes quasi-static electrical noise coupling in via the exchange interaction and via the Zeeman energy difference between the two qubits. From the fit we extract the fluctuation amplitudes $\delta_{\epsilon} = 21$ $\mu$eV and $\delta_{E_Z} = 400$ kHz. The noise in $\epsilon$ is comparable with values extracted at fridge base temperature \cite{gungordu2018pulse}, and consistent with charge noise values extracted from current fluctuation measurements of SETs (see supplemental material section IV for further information) \cite{petit2018spin, freeman2016comparison}.

To analyze the thermal impact, we characterize the temperature dependence of $T_2^*$ for two values of exchange ($J=0.5$ MHz and $J=2.5$ MHz) and we find it to be approximately stable in the range $T$ = 0.45 K - 1.2 K. While weak dependencies of $T_2^*$ have been reported in other single-qubit experiments \cite{yang2019silicon}, we observe here that the weak temperature dependence is maintained even when the exchange interaction is set to an appreciable value where we can perform two-qubit logic. 

The origin of the electrical noise limiting $T_2^*$ can potentially come from extrinsic or intrinsic sources. Although we cannot rule out all extrinsic noise sources, we have confirmed that attenuating the transmission lines does not affect the $T_2^*$ and we thus rule out a direct impact of the waveform generator and the microwave source. When intrinsic charge noise is the dominant contribution, a simple model based on an infinite amount of two-level fluctuators (TLFs) predicts a square root dependence of the dephasing rate on the temperature \cite{paladino20141}. However, this model assumes a constant activation energy distribution of the TLFs. Deviations from this assumption have been observed in SET measurements, leading to anomalous temperature dependencies \cite{connors2019low}. The small size of quantum dots, in particular SiMOS qubits, may lead to only a few TLFs being relevant for the dephasing and these may also explain the observed weak temperature dependence (see Supplemental Material section V for more details). 

Importantly, the weak dependence of $T_2^*$ on temperature makes silicon qubits remarkably robust against temperature, enabling to execute a universal quantum gate set. The ability to operate lithographically defined qubits above one Kelvin resolves one of the key challenges toward the integration of quantum hardware and control electronics all on the same chip, promising quantum integrated circuits for large-scale quantum computation.

\section*{Acknowledgements}
We thank M. Mohammad and V. V. Dobrovitski for helpful discussions and suggestions.

L.P, H.G.J.E and M. V. are funded by a Netherlands Organization of Scientific Research (NWO) VIDI grant. Research was sponsored by the Army Research Office (ARO) and was accomplished under Grant No. W911NF- 17-1-0274. The views and conclusions contained in this document are those of the authors and should not be interpreted as representing the official policies, either expressed or implied, of the Army Research Office (ARO), or the U.S. Government. The U.S. Government is authorized to reproduce and distribute reprints for Government purposes notwithstanding any copyright notation herein. 

\clearpage
\section*{Authors contributions}
L.P. and H.G.J.E. performed the experiment. H.G.J.E. fabricated the device. W.I.L.L. contributed to the process development and N.W.H contributed to the preparation of the experiment. 
J.S.C. supervised the wafer growth. L.P. and M.R. analysed the results with input from all authors. L.M.K.V and M.V. conceived the project. L.P. and M.V. wrote the manuscript with input from all authors. M.V. acquired the funding and supervised the project.

\clearpage
\onecolumngrid
\begin{center} 
    \textbf{\large Supplemental Material}
\end{center}

\section{Readout sensitivity}

\begin{figure*}[!h]%
	\includegraphics[width=\linewidth]{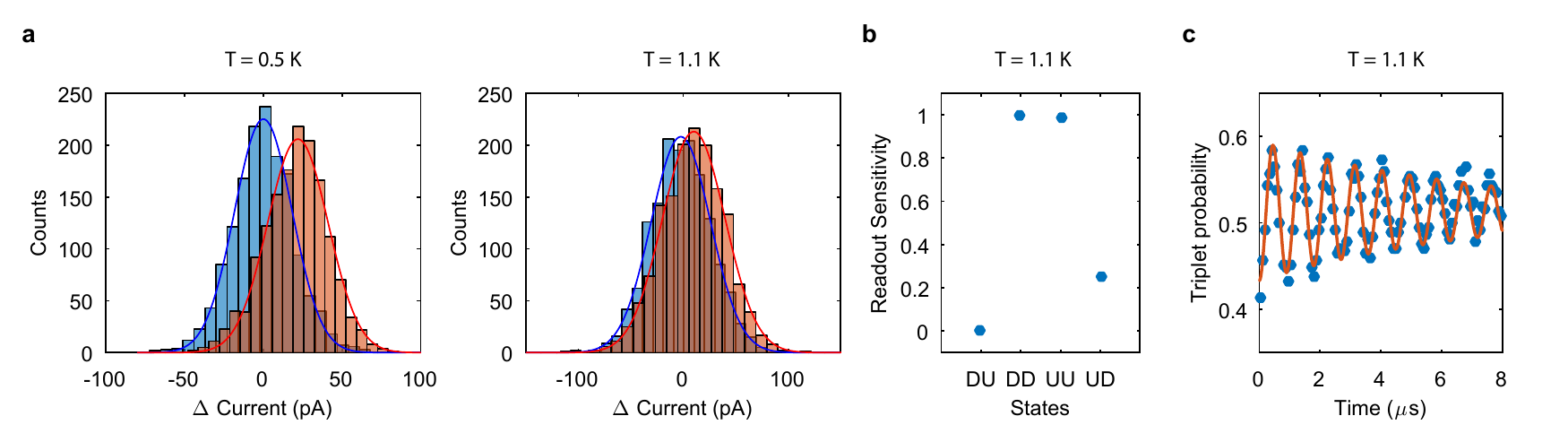}%
	\caption{\textbf{Readout characterization.}
	\textbf{a} Histograms of the readout signal for the singlet and triplet $T_-$ state for two operating temperatures. The sensitivity is reduced at higher temperatures because of the thermal broadening of the Coulomb peaks. The readout signal is obtained by subtracting a reference line, obtained from a sequence where no pulse is applied. The integration time corresponds to 40 $\mu$s.
	\textbf{b} Normalized readout signal for the four two-electron spin states. The probability of tunneling of the triplet antiparallel spin state to the (4,2) charge state is significantly lower than the tunnel rate corresponding to the two parallel spin states. 
	\textbf{c} Rabi oscillations of Q1 (see also Fig. \ref{fig:SingleQubits}b), obtained by assigning to each single-shot trace the state spin up or spin down, using a threshold obtained from the histograms in \textbf{a}. Since we subtract from the raw data a reference signal corresponding to the readout of the initialization state, the spin fraction is automatically centered around the triplet probability $P_{triplet}$ = 0.5. Nonetheless, from the data we can extract the visibility, which we find to be $V$ = 0.2 at $T$ = 1.1 K. We have normalized the triplet probability to remove the readout errors.  
	}
\label{fig:histograms}
\end{figure*}

\section{Exchange Interaction}

\begin{figure*}[!h]%
	\includegraphics[width=\linewidth]{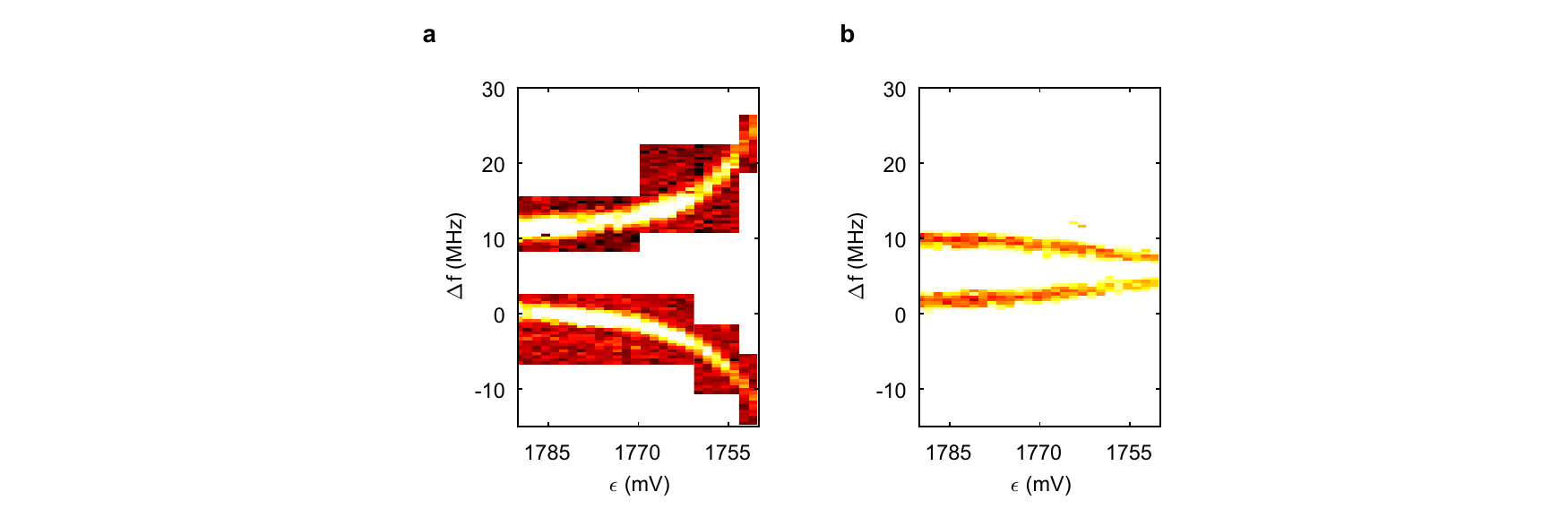}%
	\caption{\textbf{Exchange interaction.}
	\textbf{a-b} Resonance frequency of both qubits as a function of the detuning energy. In \textbf{a} we show the transitions $f_1$ and $f_4$, while in \textbf{b} we show the transitions $f_2$ and $f_3$. We measure the excited states by ESR controlled spin flips applied to the control qubit.}
\label{fig:exchange_raw}
\end{figure*}

\section{Fit of the full exchange spectrum}

We operate our device in the (5,1) charge occupation regime, close to the (4,2) transition. Considering that the four lowest-energy electrons in the right dot completely freeze out the s-orbital, they do not contribute to the dynamics and thus can be neglected. In the (5,1) charge occupation regime there are four relevant spin states in the lowest valley, $\ket{\uparrow\uparrow},\ket{\uparrow\downarrow},\ket{\downarrow\uparrow},\ket{\downarrow\downarrow}$ which are coupled via tunneling to the (4,2) singlet states $\ket{s(4,2)}$. Therefore, our system can be well-described by the Hubbard Hamiltonian~\cite{burkard1999coupled}
\begin{align}
H=\left(
\begin{array}{ccccc}
E_z & E^*_{x,2}/2 & E^*_{x,1}/2 & 0 & 0 \\
E_{x,2}/2 & \Delta E_z/2 & 0 & E^*_{x,2}/2 & t^0 \\
E_{x,1}/2 & 0 & -\Delta E_z/2 & E^*_{x,1}/2 & -t^0 \\
0 & E_{x,2}/2 & E_{x,1}/2 & -E_z & 0 \\
0 & t^0 & -t^0 & 0 &  \varepsilon - U \\
\end{array}
\right),
\end{align}
where $E_z=\mu_B(g_1 B_{z,1}+g_2 B_{z})/2$, $\Delta E_z=\mu_B(g_1 B_{z,1}-g_2 B_{z,2}$, and $E_{x,i}=\mu_B g_i (B_{x,i}+i B_{y,i})$ with $\boldsymbol{B}_i=(B_{x,i},B_{y,i},B_{z,i})^T$ and $g_i$ being the magnetic field and g-factor of quantum dot $i=1,2$. Furthermore, $t^0$ describes the tunnel coupling between the dots, $\varepsilon$ is defined as the difference of chemical potentials in quantum dot 1 and 2, and $U$ describes the charging energy of quantum dot 1. The microwave antenna allows us to apply local time-dependent transverse magnetic fields, $E_{x,i}=E_{x,i} + E^{\text{ac}}_{x,i} \cos(2\pi f^{\text{ac}} t+\phi)$, giving rise to electron spin resonance. For large detuning $|\epsilon-U|\gg t^0$ the system can be approximated by a Heisenberg Hamiltonian~\cite{burkard1999coupled}
\begin{align}
	H=J(\boldsymbol{S}_1\cdot\boldsymbol{S}_2-1/4)+\mu_B g_1\boldsymbol{B}_1\cdot\boldsymbol{S}_1+\mu_B g_2\boldsymbol{B}_2\cdot\boldsymbol{S}_2
\end{align}
with the spin operators $\boldsymbol{S}=(S_x,S_y,S_z)^T$ and the exchange interaction
\begin{align}
	J\approx 2t_0^2/(U-\varepsilon).
\end{align}
The four resonances observed in the experiment,$f_1$ ($\ket{\downarrow \uparrow} \longrightarrow \ket{\downarrow \downarrow}$), $f_2$ ($\ket{\uparrow \uparrow} \longrightarrow \ket{\uparrow \downarrow}$), $f_3$ ($\ket{\downarrow \downarrow} \longrightarrow \ket{\uparrow \downarrow}$) and $f_4$ ($\ket{\downarrow \uparrow} \longrightarrow \ket{\uparrow \uparrow}$), are then approximately given by
\begin{align}
	hf_1&=E_z-\sqrt{J^2+\Delta E_z^2}/2-J/2\\
	hf_2&=E_z-\sqrt{J^2+\Delta E_z^2}/2+J/2\\
	hf_3&=E_z+\sqrt{J^2+\Delta E_z^2}/2-/2\\
	hf_4&=E_z+\sqrt{J^2+\Delta E_z^2}/2+/2.
\end{align}
In order to extract parameters from the experiment we first fit $f_2-f_1=f_4-f_3=J/h$ to extract the exchange interaction as a function of the detuning energy and tunneling coupling. In a second step we fit $f_4+f_3-(f_2+f_1)= 2\sqrt{J^2+\Delta E_z^2}/h$. Our best fits yields $t_c=0.8\pm 0.05$ GHz and $\delta E_z = 9.1 \pm 0.1$ MHz.

\section{Noise model and noise fitting}

\begin{figure*}[!h]%
	\includegraphics[width=\linewidth]{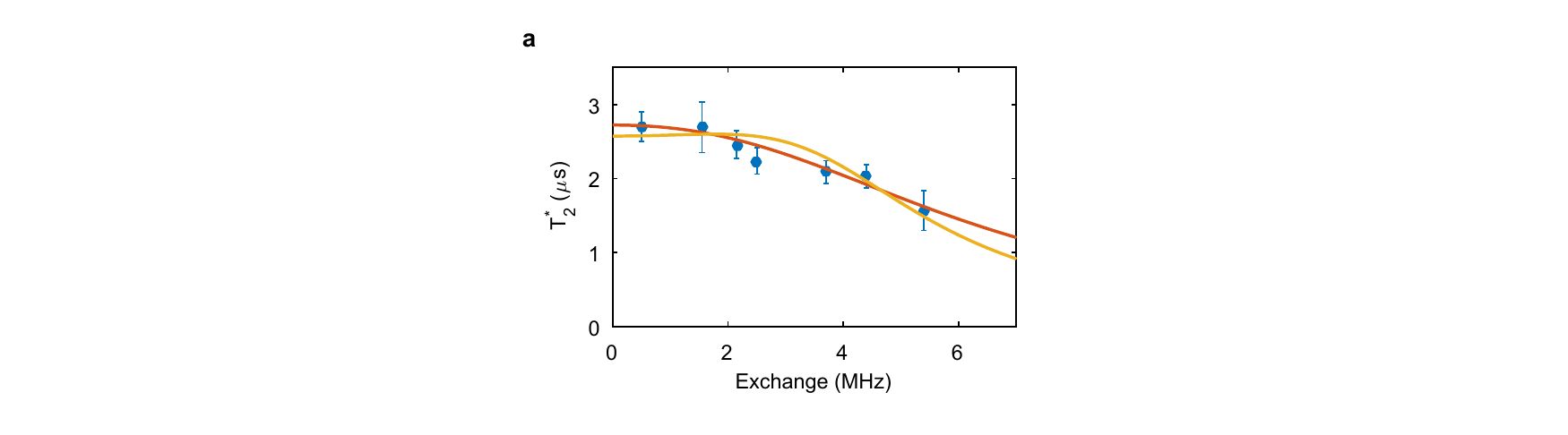}%
	\caption{\textbf{Dependence of the dephasing times on the exchange interaction.}
	\textbf{a} Same data as shown in Fig. \ref{fig:temp}a of the main text, with fits accounting for fully correlated (red line) and fully uncorrelated (yellow line) noise sources. 
	}
\label{fig:dephasing_fits}
\end{figure*}

In this section we derive analytical expressions for the pure dephasing time $T_2^*$ which are then used to extract the noise amplitude. In a Ramsey experiment the qubits are initialized in the $\ket{\downarrow\uparrow}$ state, then brought into a superposition state different for the two qubits: $\ket{Q1+}=(\ket{\downarrow\uparrow}+\ket{\uparrow\uparrow})/\sqrt{2}$ and $\ket{Q2+}=(\ket{\downarrow\uparrow}+\ket{\downarrow\downarrow})/\sqrt{2}$, then mapped back after some waiting time, and finally measured. The phase $e^{- 2\pi i \Phi(\tau)}$ the two qubits acquire during their free evolution is given by their energy difference $\Phi(\tau)=\int_0^\tau dt hf_4$ and  $\Phi(\tau)=\int_0^\tau dt f_1$. We assume that there are two dominating channels where noise can couple to the qubits, which is assumed to be longitudinally. We assume the noise comes from electrostatic fluctuations, which can couple via the detuning energy through the exchange interaction and through the g-factor modulation and spin-orbit coupling to the difference in Zeeman energy fields. The acquired phase for the two qubits are then given by
\begin{align}
	\Phi_{Q1,(Q2)}(t)=  f_{4,(1)}+ [\mathcal{D}_\epsilon \delta\epsilon(t)+\mathcal{D}_{\Delta E_z} \delta\Delta E_z(t)],
\end{align}
where $\mathcal{D}_\epsilon= \partial f_{4}/\partial J \times \partial J/\partial \epsilon $ and $\mathcal{D}_{\Delta E_z}= \partial f_{4}/\partial \Delta E_z $. 
The envelope of the Ramsey experiment is then given by the free induction decay~\cite{ithier2005decoherence}
\begin{align}
	2f(\tau)&=1+\exp\left[-\pi i \braket{\Phi(\tau)^2}\right]\\
	&1+\exp\left[-\pi t^2 \left(\mathcal{D}_\epsilon^2 \sigma_\epsilon^2 + \mathcal{D}^2_{\Delta E_z} \sigma^2_{\Delta E_z}+\kappa^2 \mathcal{D}_{\epsilon}\mathcal{D}_{\Delta E_z} \sigma_{\epsilon}\sigma_{\Delta E_z}\right)\right]
\end{align}
under the assumption of Gaussian distributed noise and zero mean $\braket{\Phi(\tau)}=0$. For the second line we assumed quasi-static noise with dispersion $\sigma_{\epsilon}=\int_{-\infty}^\infty S_\epsilon(\omega)d\omega$, where $S(\omega)=\int_{-\infty}^\infty\braket{\delta\epsilon(t)^2}e^{-i \omega t} dt$ is the power spectral density of the noise. Similar expressions hold for $\sigma_{\Delta E_z}$. The correlation coefficient is defined $\kappa=\int_{-\infty}^\infty K_{\epsilon,\Delta E_z}(\omega)d\omega/(\sigma_{\epsilon}\sigma_{\Delta E_z})$ with the cross-spectral density $K_{\epsilon,\Delta E_z}=\int_{-\infty}^\infty\braket{\delta\epsilon(t)\delta\Delta E_z(t)}e^{-i \omega t} dt$. The dephasing time is then given by 
\begin{align}
	\left(T_2^*\right)^{-1}=\sqrt{\pi} \sqrt{\mathcal{D}_\epsilon^2 \sigma_\epsilon^2 + \mathcal{D}^2_{\Delta E_z} \sigma^2_{\Delta E_z}+\kappa^2 \mathcal{D}_{\epsilon}\mathcal{D}_{\Delta E_z} \sigma_{\epsilon}\sigma_{\Delta E_z}}.
\end{align}

We fit the dephasing times as a function of exchange with either a fully uncorrelated noise $\kappa=0$ or a fully correlated $\kappa=1$ ansatz. The fits can be seen in Fig. \ref{fig:dephasing_fits}a. Our best fit yields $\sigma_{\epsilon}=21\,\mu\text{eV}$, $\sigma_{\Delta E_z}=400\,\text{kHz}$. Assuming that the origin of the noise is $1/f$ and knowing our measurement time, we can convert $\sigma_{\epsilon}$ to the value of its relative power spectrum at 1 Hz, a metric often reported in literature. We obtain $A_{\epsilon} \approx 6$ $\mu$eV$/ \sqrt{\text{Hz}}$ with $\sigma_\epsilon^2 =A_\epsilon^2 \log(f_\text{uv}/f_\text{rf}) $~\cite{ithier2005decoherence}. The lower and higher cutoff $f_\text{rf}\sim10^{-2}\, \text{Hz}$ and $f_\text{uv}\sim 10^{3}\, \text{Hz}$ are set by the experiment in the quasistatic approximation.

\section{Temperature dependence of the dephasing times}

In our experiment we observe a weak temperature dependence of $T_2^*$. This is consistent with the assumption of a few random telegraphic noise fluctuators (RTNFs). The power spectral density of a single RTNF is given by~\cite{paladino20141}
\begin{align}
	S(\omega,\nu)=\frac{A}{2 \cosh^2[E/(2k_B T)]}\frac{\nu}{\nu^2+\omega^2}.
	\label{eq:RTN}
\end{align}
Here, $A$ is the coupling strength of the fluctuations, $E$ the (activation) energy gap between the two states of the RTNF, and $\nu$ the switching rate. An explanation for the weak temperature dependence of $T_2^*$ arises from the fact that Eq.~\ref{eq:RTN} saturates if $k_B T \gg E$. Assuming that only a few RTNFs couple to our system there is only a small probability to find a RTNF which has an activation energy $E$ exactly in the temperature range between $0.4\,$K till $1.2\,$K. The same arguments hold if instead of a two-level fluctuator an Anderson impurity is the origin of charge noise \cite{beaudoin2015microscopic}.

On the other hand, if we assume a large ensemble of RTNFs a linear temperature dependence~\cite{petit2018spin} is expected assuming that the number of "activated" RTNFs increases with temperature. For a large ensemble the noise spectral density reads~\cite{paladino20141}
\begin{align}
	S(\omega)\propto \int_{2\pi f_\text{rf}}^{2\pi f_\text{uv}} \mathcal{P}(\nu,T)S(\omega,\nu)\, d\nu,	\label{eq:ensemble}
\end{align}
where $\mathcal{P}(\nu,T)$ describes the contribution of the process with a switching rate between $\nu$ and $\nu+d\nu$, thus, the probability density of a RTNF that contributes to the dephasing process. Assuming a temperature dependent switching rate $\nu = \nu_0 e^{-E/(k_B T)}$ leads to $\mathcal{P}(\nu,T)=\mathcal{P}(E,T)|\partial \nu/\partial E|^{-1}$ and one finds $\mathcal{P}(\nu,T)= \mathcal{P}(E,T) k_B T/\nu$~\cite{paladino20141}. 
Assuming further a constant distribution of activation energies, $\mathcal{P}(E,T)=\text{const}$ and inserting this into Eq.~\ref{eq:ensemble} the characteristic $1/f$ noise with a linear temperature dependence can be reproduced~\cite{gungordu2018pulse}, 
\begin{align}
    S(\omega)\approx \mathcal{P}(E,T) k_B T 
      \frac{2\pi}{\omega} \equiv\frac{A_{\epsilon}}{\omega}
\end{align}
for $2\pi f_\text{rf}\leq \omega\leq 2\pi f_\text{uv}$.
However, recent works show that the assumption of a constant distribution of activation energies $\mathcal{P}(E,T)$ is not entirely valid~\cite{connors2019low}, and this can lead to anomalous temperature dependencies.

\end{document}